\documentclass[aps,prl,twocolumn,showpacs,superscriptaddress,groupedaddress]{revtex4-1}
\usepackage{graphicx}
\usepackage{amssymb}
\usepackage{amsmath}
\usepackage{latexsym}
\usepackage{ulem}
\usepackage{color}
\usepackage{dcolumn}
\usepackage{subfigure}
\usepackage[T1]{fontenc}
\usepackage[utf8]{inputenc}
\usepackage[english,frenchb]{babel}
\usepackage[colorlinks=true,linkcolor=blue,citecolor=blue]{hyperref} %
\usepackage{bm}        % for math
\usepackage{amssymb}   % for math
\begin{document}

\title{Quantum transport in flat bands and super-metallicity}

\author{G. Bouzerar}
\email[E-mail:]{georges.bouzerar@univ-lyon1.fr}
\affiliation{CNRS, Universit\'e Claude Bernard Lyon 1, F-69622 Lyon, France}
\author{D. Mayou}
\affiliation{Université Grenoble Alpes, CNRS, Institut NEEL, F-38042 Grenoble, France}                    
\date{\today}
%\clearpage
\selectlanguage{english}
\begin{abstract}
Quantum physics in flat-band (FB) systems embodies a variety of exotic phenomenon and even counter intuitive features. The quantum transport in several graphene based compounds that exhibit a flat band and a tunable gap is investigated. Despite the localized nature of the FB states and a zero group velocity, a super-metallic (SM) phase at the FB energy is revealed. The SM phase is robust against the inelastic scattering strength and controlled only by the inter-band transitions between the FB and the dispersive bands. The SM phase appears insensitive and quasi independent of the gap amplitude and nature of the lattice (disordered or nano-patterned).  {\color{black} The universal nature of the unconventional FB transport is illustrated with the case of electrons in the Lieb lattice.}
\end{abstract}
\maketitle

Over the past decade, we are witnessing a growing interest for the physics in flat-band (FB) systems.
In these systems, and because of destructive quantum interferences the electron group velocity is exactly zero, the kinetic energy is quenched. This gives rise to various exotic physical phenomena, such as topological states \cite{tang,sun,neupert}, superconductivity \cite{miyahara,cao}, Wigner crystal \cite{wu1,wu2} and ferromagnetism \cite{tasaki,mielke,noda}. The wealth and fascinating physics that take place in these systems motivate the search for efficient procedures and strategies for flat-band engineering. For instance, twisted bilayer graphene is known to feature isolated and relatively flat bands near charge neutrality, when tuned to special magic angles only \cite{dossantos,morell,macdo,trambly1}. Recently, it has been suggested that robust FB can be realized in van der Waals patterned dielectric superlattices that could be controlled by gate voltage\cite{likun}.
Nanolithography, molecular engineering and 3D printing are also possible pathways to design complex two dimensional materials \cite{fowlkes,vyatskikh,kempkes,berenschot}. The field of cold atoms on artificial lattices also offers a plateform to address these fundamental issues  since it allows the direct tuning of the physical parameters of the model Hamiltonians \cite{belopolski,bloch,lewenstein,cooper}.

The important progress made in the realization of complex and nanostructured materials has stimulated theoretical studies in fractalized systems \cite{vanveen1,vanveen2,fremling,westerhout,brzezinska}. Recently, considering the case of the graphene Sierpinski carpet where the fractalization induces (i) a $E=0$ flat band and (ii) a gap in the spectrum, we have reported an unusual form of quantum electronic transport \cite{bouzerar-mayou}. Despite the gap, an unexpected super-metallic (SM) phase, insensitive to the strength of the inelastic scattering rate appears at the neutrality point with a conductivity that coincides within few percent with $\sigma_0 =\frac{4e^2}{\pi h}$ that of the pristine compound. In this system, the transport is controlled by inter-band transitions only, between the FB and the valence (conduction) band.  

Our goal is to address the crucial and inevitable question that naturally rises: Is this unusual form of quantum electronic transport universal? More precisely, does the SM flat-band transport take place in other type of systems? For that purpose, we consider three different situations that lead to a FB at $E=0$ and a gap in the spectrum: (i) the fully uncompensated graphene (FUG) where vacancies are randomly distributed on the same sub-lattice and two self-similar lattices, (ii) the Serpinski carpet (GSC) and (iii) the Sierpinski gasket (GSG). 
The choice for graphene is also motivated by the fact that it has emerged as an outstanding system for fundamental research \cite{das-sarma,castro-neto,geim,falko}. Note that the GSC conductivity as studied in details in Ref.\cite{bouzerar-mayou}, will be used just for comparison with the gasket case. Transport is expected to be drastically different in the gasket than in the carpet. The Sierpinski carpet is infinitely ramified while the gasket only finitely. In other words, the gasket can be deconstructed by removing a finite number of sites while it requires an infinite one for the carpet \cite{topology}. 

To address the second question, we consider the electronic transport in the Lieb lattice (the CuO$_2$ planes in cuprates)  where the spectrum is gapless and a FB meets the conduction band and the valence band at the Dirac point. It is nowadays possible to realize experimentally the Lieb lattice either by manipulating cold atoms in optical lattices \cite{shen,goldman,apaja}, or by direct laser writing of optical waveguides \cite{vicencio,guzman,mukherjee}, and it could be even synthesized by means of covalent organic frameworks \cite{cui}. 

\begin{figure}[t]\centerline
{\includegraphics[width=0.9\columnwidth,angle=0]{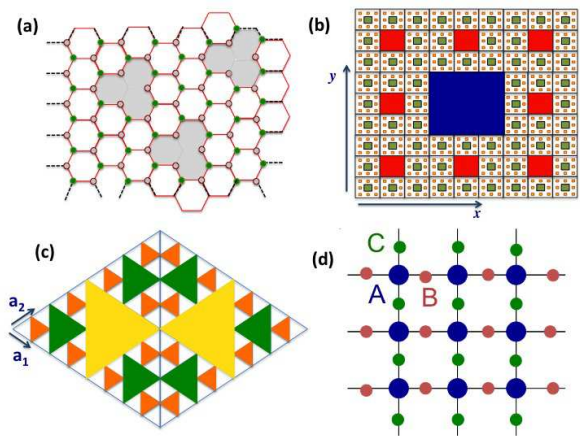}}
\caption{(Color online) Illustration of (a) the fully uncompensated graphene (FUG), (b,c) the graphene Sierpinski carpet (GSC) and gasket (GSG) and (d) the Lieb lattice. In (a), (b) and (c) the coloured area correspond to the regions of removed atoms.
}
\label{fig1}
\end{figure} 

Electrons in the FB systems, as illustrated in Fig.~\ref{fig1}, are modeled by a nearest-neighbour tight-binding Hamiltonian that reads,
\begin{eqnarray}
\widehat{H}=-t \sum_{\left\langle ij\right\rangle,s} c_{is}^{\dagger}c^{}_{js} +h.c.,
\end{eqnarray}
$\left\langle ij\right\rangle$ denotes nearest neighbor pairs. c$_{is}^{\dagger}$ creates an electron with spin $s$ at site \textbf{R}$_{i}$. In the Lieb lattice the only allowed hoppings are between the nearest neighbor pairs (A,B) and (A,C).

The GSC is constructed from a square piece of graphene of length $L = 3^{i_{c}+1} a$ ($a$ is the nearest neighbour C-C distance). We use for the GSC's the notation ($i_{c}$,$f$) where $f$ is the degree of "fractalization" that varies from 0 (pristine) to its maximum value $f_{max}=i_{c}$.  The GSG is obtained from a triangular piece of graphene delimited by the vectors $N.{\bf a }_1$ and $N.{\bf a }_2$ where ${\bf a }_1$ and ${\bf a }_2$ are the unit cell vectors of graphene and $N=2^{i_{g}+1}$. It is then symmetrized with respect to the $y$-axis to give a diamond piece of graphene. Because of the symmetrization, the GSG contains the same number of C atoms on both sub-lattices. The GSG is specified by the notation ($i_g$, $f$). Here, our study is restricted to optimally fractalized compounds only: $f=i_{c}$ for the carpet and $f=i_{g}$ for the gasket. The  lattice geometry is unimportant for the FUG case. Periodic boundary conditions along $x$ and $y$ directions (see Fig.~\ref{fig1}) are used for the FUG, the GSC and the Lieb lattice and along ${\bf a }_1$ and ${\bf a }_2$ for the GSG. 

\begin{figure}[t]\centerline
{\includegraphics[width=0.9\columnwidth,angle=0]{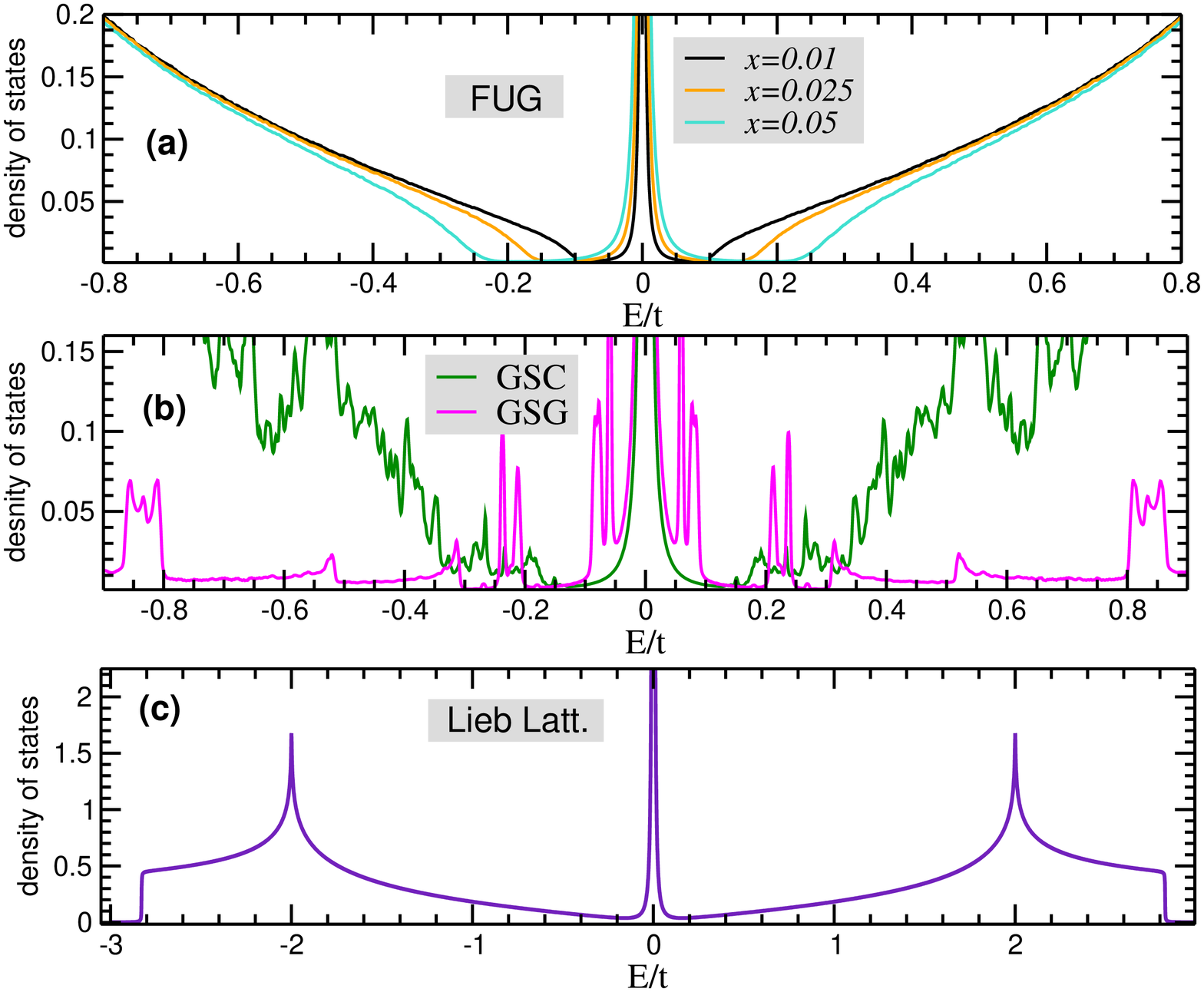}}
%{Fig2-new.eps}}
\caption{(Color online) Density of states (in $1/t$) in (a) the FUG for three different concentrations of vacancies, $x=0.01$, $0.025$ and $0.05$, (b) in the GSG and in the GSC and (c) in the Lieb lattice. The systems used for the calculations are (7,7) for the GSC, (11,11) for the GSG (see the notations in the text). The FUG contains approximately $3.5\times 10^6$ sites.}
\label{fig2}
\end{figure} 

The conductivity along the $x$-direction is given by the Kubo formula,
\begin{eqnarray}
\sigma(E)=\frac{e^{2}\hbar}{\pi\Omega} \mathrm{Tr} \left[\operatorname{Im} \widehat{G}(E) 
 \widehat{v}_{x} \operatorname{Im} \widehat{G}(E) \widehat{v}_{x} \right]. 
\label{eqcond}
\end{eqnarray}
The current operator is defined by $ \widehat{v}_{x} =  -\frac{i}{\hbar}\left[ \widehat{x} ,\widehat{H} \right]$ and the Green's function $\widehat{G}(E)=(E+i\eta-\widehat{H})^{-1}$. $\Omega$ is the sample area and $\eta$ mimics an energy independent inelastic scattering rate with a characteristic timescale $\tau_{in}= \frac{\hbar}{\eta}$. For the FUG, the GSG and the GSC the calculations are done using the Chebyshev polynomial Green's function method (CPGF) \cite{mucciolo,weisse} that (i) allows large scale calculations as it requires a modest amount of memory and (ii) a CPU cost that varies only linearly with the system size $N_S$. CPGF has proven to be a powerful tool to address the nature of the magnetic couplings in disordered materials \cite{richard,lee}. In the same spirit as CPGF, the conductivity could be also calculated by quantum wave packet dynamics as well \cite{trambly,cresti,triozon}.
The FUG, the GSC and the GSG considered contain approximately $3.5\times 10^6$ sites. The number of random vectors $N_{R}$ used for the stochastic trace calculation is 50. The number of Chebyshev polynomials kept is $M=2 500$, leading to a $M \times M$ matrix of moments used for the conductivity calculation. It has been checked that both $N_{R}$ and $M$ were sufficient to reach convergence. On the other hand, the calculations are realized analytically in the case of the Lieb lattice.
\begin{figure*}
 \includegraphics[width=1.80\columnwidth,height=6cm]{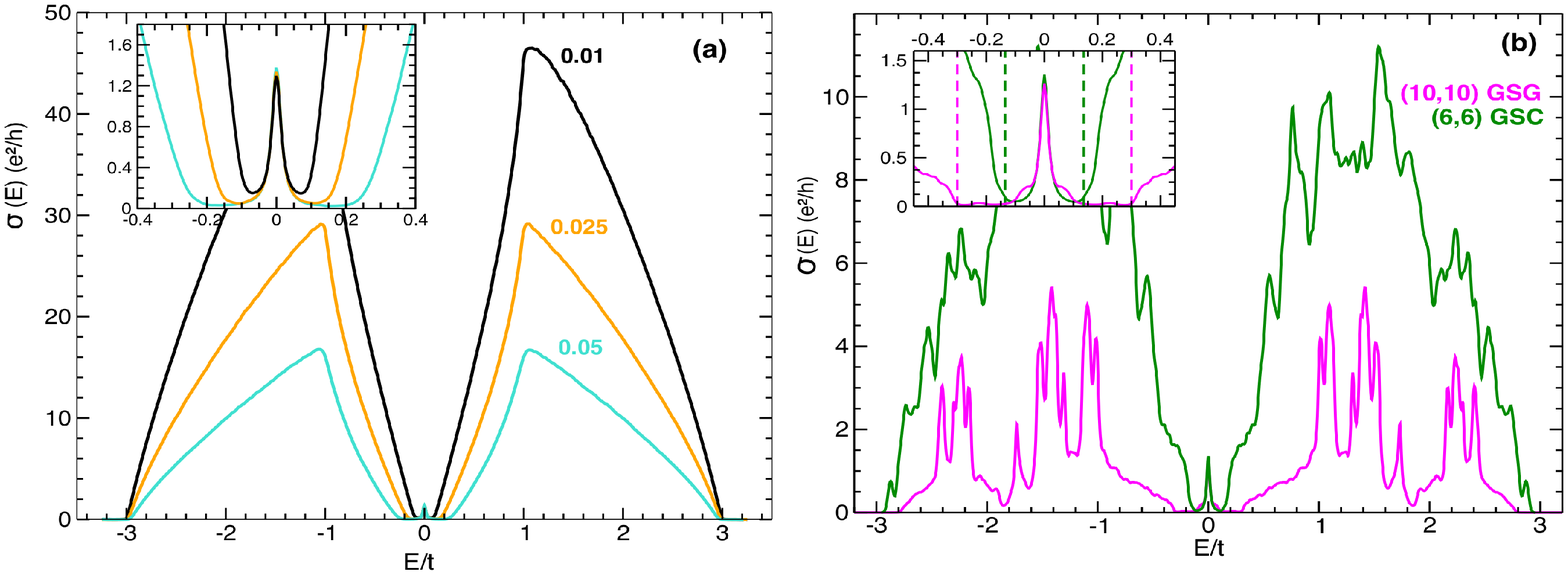}
 \caption{(Color online) Conductivity (in $\frac{e^2}{h}$) at T=0~K as a function of the energy in (a) the FUG and (b) the GSC and GSG. In (a) we consider three different concentrations of vacancies $x=$ 0.01, 0.025 and 0.05, the systems contain approximately $3.5\times 10^6$ sites.
In (b) the GSC and the GSG are respectively (6, 6) and (10, 10) systems. $\sigma(0)$ is $1.26 \frac{e^{2}}{h}$ in the GSG and $1.365 \frac{e^{2}}{h}$ in the GSC. Here $\eta =0.016~t$ but $\sigma(0)$ is found insensitive to $\eta$. The insets magnify the neutrality point region.
}
\label{fig3ab}
\end{figure*}

Figure~\ref{fig2} depicts the electronic density of states (DOS) $\rho(E)=-\frac{1}{\pi N_S} \mathrm{Tr} \left[ \operatorname{Im} \widehat{G}(E)\right]$ as a function of the energy in the graphene compounds and in the Lieb lattice. As expected for the FUG, a $\delta$-peak at $E=0$ appears and a gap which increases with the density of randomly distributed A-vacancies \cite{pereirab}. The gap ($\Delta$) from the valence (resp. conduction) band to the FB of zero energy modes (ZEM) is $0.10 ~t$, $0.15~ t$ and $0.20 ~t$, for respectively, $x= 0.01$, $0.025$ and $0.05$. The DOS has a richer texture in the fractal lattices. Besides a gap, $\Delta=0.135 ~t$ in the GSC and significantly larger in the GSG where it is approximately $0.31~t$, we observe complex fluctuating sub-structures that result from the fractal nature of the eigenspectrum. In the GSG, we observe many extended low-DOS regions interspersed by sharp peaks. This reflects a one-dimensional-like characteristic that originates from the finitely ramified fractal lattice.
In the GSG, in addition to a central ZEM peak, several pronounced satellite peaks appear at $E=\pm 0.06~t$, $\pm 0.077~t$,  $\pm 0.085~t$, $\pm 0.22~t$ and $\pm 0.24~t$, revealing additional, almost flat bands. The exact diagonalisation calculations on smaller systems, (4,4), (5,5) and (6,6) have confirmed that these sub-bands are not rigorously flat, in contrast to the $E=0$ band.
In addition, in both the FUG and the GSC the number of ZEM states ($N _{ZEM}$) is exactly $\vert N_{A}-N_{B}\vert$, N$_{A}$ (resp. N$_{B}$) being the number of C atoms on sublattice A (resp. B), as it is expected in bipartite lattices \cite{lieb,pereirab}. In the GSC, the ZEM density, $x_{ZEM}$, is approximately 0.05. In contrast, the situation is different in the GSG, where by construction $N_{A}$=$N_{B}$ (see Fig.1). The expected $x_{ZEM}$ should be zero, which is not the case. It varies with the system size and we find $x_{ZEM}=0.164$, 0.172, 0.177 and 0.178 in the (8,8), (9,9), (10,10) and (11,11) respectively, indicating a convergence towards 0.18. If $N^{L} _{A}$ (resp. $N^{L} _{B}$) is the number of A (resp. B) sites of the 'left' triangle of the GSG diamond, $\vert N^{L} _{A} - N^{L} _{B} \vert$ is also different from $N _{ZEM}$. 
Fig.~\ref{fig2}(c) illustrates the well known DOS in the Lieb lattice. It reveals 3 bands, two dispersive, which form a Dirac cone at the M point of the  Brillouin zone and a FB at $E=0$. We recall, in this case, that the local charge density of the localized, $E=0$ states, is non zero on B and C-sublattices only.

We discuss the electronic transport in these systems, with a focus on the central region. In the FUG, the conductivity, $\sigma(E)$, is depicted in Fig.~\ref{fig3ab}(a) for different concentrations of vacancies. Besides a maximum in the valence band (VB) and conduction band (CB) at $E=\pm t$ (Van Hove singularities in pristine graphene), $\sigma(E)$ is finite for $\vert E \vert \geq \Delta$ and decreases, as expected, as $x$ increases. However, a close look at the FB vicinity reveals a peak that varies very weakly with $x$, $\sigma (0)$ coincides within few percent with that of the pristine case, $\sigma_0$. We have also checked that $\sigma (0)$ is insensitive to $\eta$, with $\eta$ ranging from $0.001~t$ to $0.05~t$. We should stress that our calculations correspond to the thermodynamic limit, as it is illustrated in the supplementary material (SM) part (see below). For $\vert E \vert \leq \Delta$, $\sigma(E)$ gets narrower and narrower as $\eta$ decreases and can be nicely fitted by a Lorentzian of width $\eta$. However, our results disagree with those of Ref. \cite{cresti} where $\sigma(0)=0$ is found. In this work, $\sigma(E)$ is obtained from the Einstein formula and a direct calculation of the diffusivity from wave packet propagation. The singular DOS at $E=0$ and the fact that their calculations correspond to the limit $\eta=0$, may explain the discrepancy.

Let us consider how self-similarity affects the electronic transport. Results, for a fixed $\eta$, are depicted in Fig.~\ref{fig3ab}(b). 
The conductivity in the GSC has been discussed in details in Ref.\cite{bouzerar-mayou}. It is only considered to facilitate the comparison with the gasket case and show the universality of the FB quantum transport. In the GSG, $\sigma(E)$ is much smaller than that of the GSC and the peaks appear sharper. In the inset, we observe a clear gap in the GSG of $0.31~t$ much larger that that of the GSC ($0.135~t$), as seen in the DOS (Fig.~\ref{fig2}). A peak at $E=0$ is also clearly visible with values close to $\sigma_0$. More precisely, we find  $\sigma(0)= 1.07 \sigma_0$ in the GSC and $0.99 \sigma_0$ in the GSG. Note also, for the GSG, shoulders in the central peak, that are absent for the GSC. They correspond to the states located at $E=\pm 0.06~t$, $\pm 0.077~t$ and $\pm 0.085~t$ in the DOS. We have checked that, these shoulders disappear as $\eta$ reduces (see the supplementary material  part). Compared to the FB states, these satellite states behave in a more "standard" way. They are localized impurity states, leading to a vanishing conductivity when $\eta \rightarrow 0$. These results are robust, with negligible size effects (see the the supplementary material part). Hence, from Fig.\ref{fig3ab}(a) and Fig.\ref{fig3ab}(b) we conclude that these graphene based systems lead to the same conclusion: a universal quantum transport at $E=0$ with a super-metallic flat band and a conductivity that reduces to the inter-band term (the intra-band contribution vanishes). Remark that an important inter-band term was also at the origin of the quantum electronic transport anomalies in the icosahedral quasicrystals $\alpha$-AlMnSi \cite{trambly2}.
The inter-band super-metallic regime can be visualized as a quantum transport controlled by the velocity fluctuations in systems where its average is very small or zero.

\begin{figure}[t]\centerline
%{\includegraphics[width=\columnwidth,angle=0]{fig4new.eps}}
{\includegraphics[width=0.95\columnwidth,angle=0]{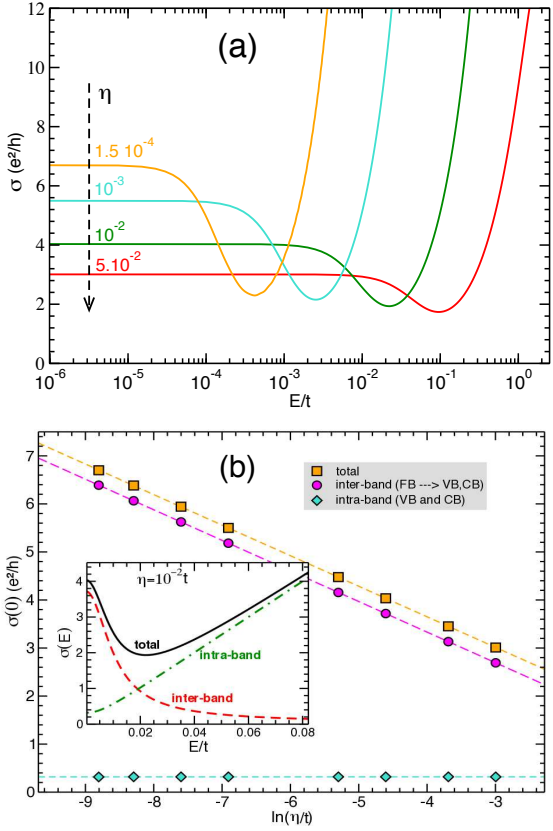}}
\caption{(Color online) 
(a) Conductivity in the Lieb lattice as a function of the energy $E$ for different values of $\eta$. (b) Different contributions to $\sigma(0)$ as a function of $ln(\eta/t)$.(Inset) $\sigma(E)$ (total, inter-band and intra-band) as a function of E for $\eta=2.10^{-3}~t$. 
} 
\label{fig4}
\end{figure} 
 {\color{black} 
Finally, we address the possibility of FB induced SM phase in a very different system, the Lieb lattice. Fig.\ref{fig4} (a) depicts $\sigma(E)$ as a function of E for different values of $\eta$. As in the graphene based systems, a peak at $E=0$ is revealed (more visible in the inset of Fig.\ref{fig4} (b)). However, in the Lieb lattice, $\sigma(E=0)$ increases slowly as $\eta$ decreases ($\eta$ varies by two orders of magnitude). 
The inset of Fig.\ref{fig4} (b) shows, for $\eta/t=10^{-2}$, the decomposition in terms of the intra-band ($\sigma_{intra}$) and the inter-band ($\sigma_{inter}$) contributions. The only non vanishing matrix elements of the velocity operator that contribute to $\sigma_{inter}$, are between the FB and the CB (resp. VB) states, those between VB and CB states are zero. We find that $\sigma_{intra}$ is finite at $E=0$. A focus on the $\eta$-dependence of $\sigma(0)$, as it is plotted in Fig.\ref{fig4} (b), shows that $\sigma_{intra}(0)$ is constant and equals $0.318 \frac{e^2}{h}$. 
On the other hand, $\sigma_{inter}(0)$ has an unusual logarithmic dependence on $\eta$. We find, numerically, $\sigma_{inter}(0)= \sigma_{1}+ \sigma_{2}\vert ln(\eta/t)\vert$ where $\sigma_{1}=0.784 \frac{e^2}{h}$ and $\sigma_{2}=0.637 \frac{e^2}{h}$. Using the linear dispersion
of the dispersive bands in the vicinity of the Dirac point, we obtain the analytical expressions: $\sigma_{intra}(0)=\frac{1}{\pi} \frac{e^2}{h}$, $\sigma_{2}=\frac{2}{\pi} \frac{e^2}{h}$. $\sigma_{1}$ depends on the cut-off energy ($E_{c}$) and leads to $\sigma_{1}=\frac{1}{\pi} \frac{e^2}{h} ln(E_{c}^2/t^2)$. Using a normalized DOS for the dispersive bands gives $0.806 \frac{e^2}{h}$.

We propose now to discuss the $\eta$-dependence of the diffusivity in the SM phase. In the gapped cases, for both $\eta$ and $\vert E\vert$ smaller than $\Delta$, $\rho(E)$ reduces to $\frac{N_{ZEM}}{\pi\Omega} \frac{\eta}{E^{2}+\eta^{2}}$. Eq.(\ref{eqcond}) of the conductivity can be re-written,
\begin{eqnarray}
\sigma(E)= \left( \frac{4\hbar}{N_{zem}} \sum_{\alpha,\lambda=\pm,\beta} 
\dfrac{\vert \langle  \Psi_{\beta} \vert \widehat{v}_{x}\vert \Phi^{\lambda}_{\alpha} \rangle \vert^{2}}{E_{\alpha}^2 }\eta \right) e^{2} \rho(E), 
\end{eqnarray}
where we have introduced $\vert \Phi^{\lambda}_{\alpha} \rangle$, the valence ($\lambda=-$) and conduction ($\lambda=+$) eigenstates with energy $\pm \vert E_{\alpha} \vert$ and the FB eigenstates $\vert \Psi_{\beta} \rangle$. From the Einstein formula, the diffusivity $D(E)=\frac{\sigma(E)}{e^{2}\rho(E)}$ is straightforwardly obtained. It scales linearly with $\eta$, instead of the $1/\eta$ behaviour in standard metallic systems where $D=\frac{1}{2}v^{2}_{F}\frac{\hbar}{\eta}$. In the gapless case of the Lieb lattice, the transport is still controlled by the inter-band term but the diffusivity has now two contributions, $D=D_0 \eta + D_1 \vert \eta.ln(\eta) \vert$. We expect, by introducing vacancies in the Lieb lattice that a gap should open and the conductivity might loose the $\vert ln(\eta) \vert$ contribution and $\sigma_{intra}(0)$ should vanish. All the features reported here, justify the use of the term "super-metallicity" and generalize, what has been found in the peculiar case of the GSC \cite{bouzerar-mayou}.

In conclusion, in standard systems, the quantum transport is dictated by the average intra-band velocity of the carriers, here at the FB energy, it is of  inter-band nature. In all cases investigated, a SM phase, controlled by the off-diagonal matrix elements of the current operator, is revealed at the FB energy. In the graphene based systems, the conductivity is independent of the gap value, nature of the lattice and inelastic scattering strength, and coincides within few percent with $\sigma_0$ ($\frac{4e^{2}}{\pi h}$).
In the gapless case of electrons the Lieb lattice, the FB conductivity is found to vary logarithmically with the inelastic scattering strength ($\sigma \approx \frac{1}{2} \sigma_0 \vert ln(\eta)\vert$). This shows that the unconventional super-metallicity of the flat bands has a universal character. Based on the recent progress in the realization of complex 2D systems and in optical lattice physics, we hope that our findings will stimulate experimental studies.}

\newpage

\section*{\bf Supplementary Material.} 
            
%\date{\today}
%\clearpage
\selectlanguage{english}

We provide some additional informations that support (i) the absence of finite size effects in the data presented in our manuscript and (ii) show how the electronic conductivity, in the vicinity of the neutrality point, is affected by the inelastic scattering strength. We consider the case of the fully uncompensated graphene (FUG) and that of the graphene Sierpinski gasket (GSC). We recall that in the FUG, vacancies are randomly distributed on the same sublattice.

The calculations shown below are performed within the Kubo formalism using the Chebyshev polynomial Green's function method (CPGF) \cite{mucciolo,garcia,weisse}. The CPGF approach is a powerful method. It allows large scale real-space numerical calculations (particularly suitable for disordered systems) as it requires a modest amount of memory and a CPU cost that varies only linearly with the system size $N$. This is in contrast with the exact diagonalization calculations technique that is very demanding in memory and CPU, they scale respectively as $N^2$ and $N^3$. The CPGF has proven, for instance, to be a powerful tool to address the nature of the magnetic couplings in disordered systems \cite{richard,lee}. Note that, in the same spirit as CPGF, the conductivity could be also calculated by quantum wave packet dynamics as well \cite{trambly,cresti,triozon}.
\begin{figure}[t]\centerline
{\includegraphics[width=1.0\columnwidth,angle=0]{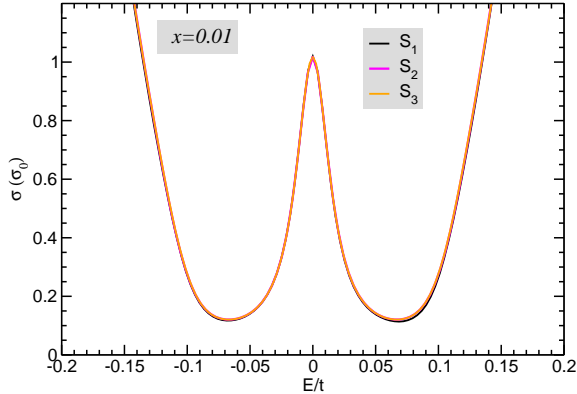}}
\caption{(Color online) 
Conductivity (in units of $\sigma_0=\frac{4e^2}{\pi h}$) as a function of $E/t$ in the FUG for 3 different systems ($S_i$, i=1, 2 and 3). The systems are square pieces of graphene with periodic boundary conditions (in both $x$ and $y$-directions).
$S_1$, $S_2$ and $S_3$ contain respectively about $N_i= 4\times10^5 $, $3.6\times10^6 $ and $33\times10^6 $ sites.
The concentration of A-type vacancies is $x=0.01$ and the inelastic scattering rate $\eta$ is set to $0.016 t$.
}
\label{figsupp1}
\end{figure} 

The conductivity along the $x$-direction is given by the Kubo-Greenwood formula \cite{kubo,greenwood},
\begin{eqnarray}
\sigma(E)=\frac{e^{2}\hbar}{\pi\Omega} \mathrm{Tr} \left[\operatorname{Im} \widehat{G}(E) 
 \widehat{v}_{x} \operatorname{Im} \widehat{G}(E) \widehat{v}_{x} \right]. 
\label{eqcond}
\end{eqnarray}
The current operator is defined by $ \widehat{v}_{x} =  -\frac{i}{\hbar}\left[ \widehat{x} ,\widehat{H} \right]$ and the Green's function $\widehat{G}(E)=(E+i\eta-\widehat{H})^{-1}$. $\Omega$ is the sample area and $\eta$ mimics an energy independent inelastic scattering rate with a characteristic timescale $\tau_{in}= \frac{\hbar}{\eta}$. 

To evaluate stochastically the trace that enters the dc conductivity expression of Eq. (\ref{eqcond}), the number of random vectors $N_{R}$ used is 200 for the smallest system and 10 for the largest. The number of Chebyshev polynomials kept is respectively, $M=$ 2 500, 4 000 and 6 000 for $\eta=0.016t$,  $\eta=0.008t$ and $\eta=0.004t$.  We recall that the calculation of the conductivity requires, for a given random vector, the determination of a $M \times M$ matrix of moments. We have systematically checked that both $N_{R}$ and $M$ were sufficient to reach convergence within less than 1-2 \% accuracy. 
\begin{figure}[t]\centerline
{\includegraphics[width=1.0\columnwidth,angle=0]{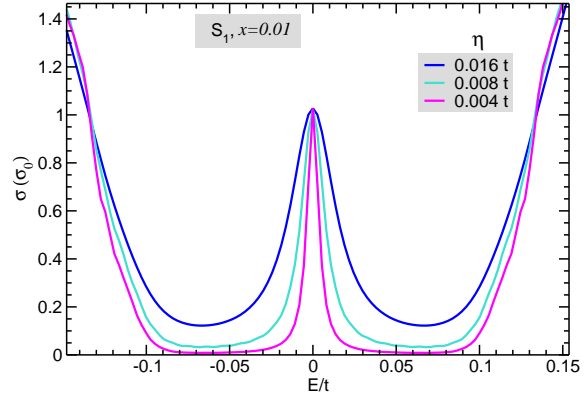}}
\caption{(Color online) 
Conductivity (in units of $\sigma_0=\frac{4e^2}{\pi h}$) in the vicinity of the flat band energy as a function of $E/t$ in the FUG for several values of the inelastic scattering rate $\eta$. The system $S_1$ contain about $4\times10^5$ C atoms. The concentration of A-type vacancies is $x=0.01$.
}
\label{figsupp2}
\end{figure} 

Fig.\ref{figsupp1} depicts the effects of the system size on the calculated dc-conductivity in the fully uncompensated graphene. Here, the concentration of vacancies is fixed and set to $x=0.01$.
We observe, for the chosen value of the inelastic scattering rate, the absence of size effects. As the system size increases the conductivity in the vicinity of the neutrality point appears essentially unaffected. The next figure, Fig.\ref{figsupp2} illustrates the effects of the inelastic scattering rate $\eta$ on the dc-conductivity in the FUG. The concentration of vacancies is again kept fixed ($x=0.01$). As $\eta$ is decreased we observe a narrowing of the peak centred at $E=0$ but $\sigma(0)$ remains unchanged. The peak can be nicely fitted by a Lorentzian of half width $\eta$. The super-metallic phase at $E=0$ is robust.
\begin{figure}[t]\centerline
{\includegraphics[width=1.0\columnwidth,angle=0]{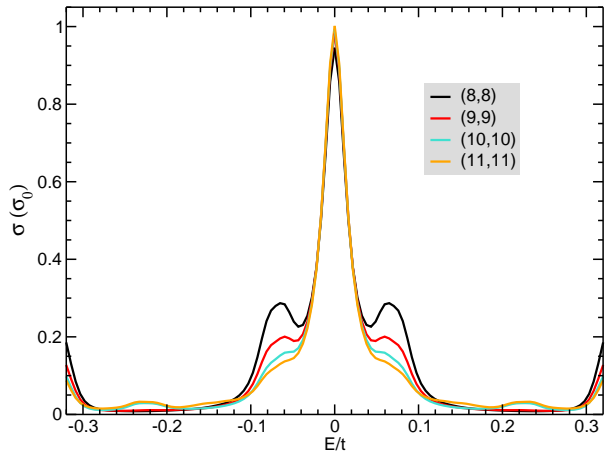}}
\caption{(Color online) 
Conductivity (in units of $\sigma_0=\frac{4e^2}{\pi h}$) as a function of $E/t$ in optimally fractalized GSG for various system sizes. The $(n,f=n)$ gaskets (as defined in the main text) contain respectively about $10^5$, $3\times10^5$, $9\times10^5$ and $2.7\times10^6$ C atoms for $n=8$, 9, 10 and 11. The inelastic scattering rate $\eta$ is fixed and set to $0.016 t$.
}
\label{figsupp3}
\end{figure} 

We now show some supplementary data that concern the graphene Sierpinski gasket (GSC). Fig.\ref{figsupp3} shows the effects of the system size on the conductivity plotted as a function of the energy (in the vicinity of the neutrality point) in the GSC. First, we observe satellite peaks at $E=\pm 0.07 ~t$ that reduces as the system size increases. We also observe at higher energy, $E=\pm 0.23 ~t$, a small peak that develops when the system size increases. This suggests a transfer of weight from the $E=\pm 0.07 ~t$ peak to the higher energy peak. Note that these peaks are absent in the case of the Sierpinski carpet. The calculated values at $E=0$ show some small size effects, $\sigma(E=0) =$ 0.95 $\sigma_0$,  0.98 $\sigma_0$, 0.99 $\sigma_0$ and 1.0 $\sigma_0$ respectively for the (8,8), (9,9), (10,10) and (11,11) GSG's, where $\sigma_0 = \frac{4}{\pi} 
\frac{e^2}{h}$ is the value in the pristine compound. The notation $(n,f)$ is defined in the main text.

\begin{figure}[t]\centerline
{\includegraphics[width=1.0\columnwidth,angle=0]{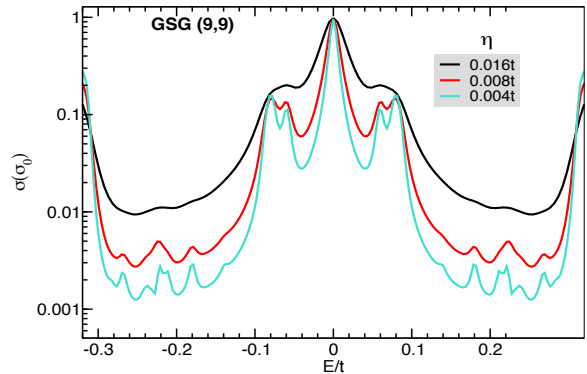}}
\caption{(Color online) 
Conductivity (in $\sigma_0=\frac{4e^2}{\pi h}$) as a function of $E/t$ (in the vicinity of the neutrality point) in the (9,9) GSG for several values of the inelastic scattering rate $\eta$ as depicted in the figure.
}
\label{figsupp4}
\end{figure} 

The effects of the inelastic scattering rate $\eta$ on the conductivity is illustrated in Fig.\ref{figsupp4} for the (9,9) GSG. As $\eta$ decreases, we see that the substructure of the peaks becomes more and more visible. In addition, we observe a strong suppression of the conductivity at the peak's energy. The reduction is much stronger for the high energy peaks. As $\eta$ reduces, the phase coherence length $L_{\phi}$ ($\propto 1/\eta$) increases, thus the quantum interference effects become more and more important, leading to Anderson localization and to a vanishing conductivity in the limit $\eta \rightarrow 0$ and $N \rightarrow \infty$. The states associated to the satellite peaks behave in a more "standard" way, as usual impurity band states. In contrast, at the neutrality point, the conductivity remains totally insensitive to $\eta$, revealing the different nature of the flat band states.

\begin{acknowledgments}
\end{acknowledgments}

\newpage


\begin{thebibliography}{99}
%\bibliography{apssamp}


\bibitem{tang}E. Tang, J.-W. Mei, and X.-G. Wen, Phys. Rev. Lett. \textbf{106}, 236802 (2011).
\bibitem{sun}K. Sun, Z. Gu, H. Katsura, and S. Das Sarma, Phys. Rev. Lett. \textbf{106}, 236803 (2011).

\bibitem{neupert}T. Neupert, L. Santos, C. Chamon, and C. Mudry, Phys. Rev. Lett. \textbf{106}, 236804 (2011).


\bibitem{miyahara}S. Miyahara, S. Kusuta, and N. Furukawa, Physica C: Superconductivity \textbf{460,} 1145 (2007).

\bibitem{cao} Y. Cao et al., Nature \textbf{556}, 43 (2018).


\bibitem{wu1}C. Wu, D. Bergman, L. Balents, and S. Das Sarma, Phys. Rev. Lett. \textbf{99}, 070401 (2007).
\bibitem{wu2}C. Wu and S. Das Sarma, Phys. Rev. B \textbf{77,} 235107 (2008).


\bibitem{tasaki}  H. Tasaki, Phys. Rev. Lett. \textbf{ 69} 1608 (1992); Prog. Theor. Phys. \textbf{ 99}(4):\textbf{489} (1998).
\bibitem{mielke}  A. Mielke, Phys. Rev. Lett. \textbf{82}, 4312 (1999).
\bibitem{noda} K. Noda, A. Koga, N. Kawakami, and T.Pruschke, Phys. Rev. A \textbf{80}, 063622 (2009)




\bibitem{dossantos} J. M. B. Lopes dos Santos, N. M. R. Peres, and A. H. Castro Neto, Phys. Rev. Lett. \textbf{99}, 256802 (2007).
\bibitem{morell} E. Suarez Morell, J. D. Correa, P. Vargas, M. Pacheco, and Z. Barticevic, Phys. Rev. B \textbf{82}, 121407 (2010).
\bibitem{macdo} R. Bistritzer and A. H. MacDonald, Proc. Natl. Acad. Sci. USA \textbf{108}, 12233 (2011).
\bibitem{trambly1} G. Trambly de Laissardiere, D. Mayou, and L. Magaud, Phys. Rev. B \textbf{86}, 125413 (2012).

\bibitem{likun}Li-kun Shi et al. 2D Mater. \textbf{7} 015028 (2020).


\bibitem{fowlkes} J. Fowlkes et al., ACS Nano {\bf 10}, 6163 (2016).
\bibitem{vyatskikh} A. Vyatskikh, Nature Comm. {\bf 9}, 593 (2018).
\bibitem{kempkes} S.N. Kempkes et al., Nature Physics {\bf 15}, 127 (2019).
\bibitem{berenschot} E.J.W. Berenschot, H.V. Jansen and N.R. Tas, J. of Micromech. Microeng.,  {\bf 23}  055024 (2013).


\bibitem{belopolski}I. Belopolski et al., Sci. Adv. {\bf 3}, e1501692 (2017).
\bibitem{bloch}I. Bloch, J. Dalibard, W. Zwenger, Rev. Mod. Phys. {\bf 80}, 885 (2008). 
\bibitem{lewenstein}M. Lewenstein et al., Adv. Phys. {\bf 56}, 243 (2007).
\bibitem{cooper}N.R. Cooper, J. Dalibard and I.B. Spielman, Rev. Mod. Phys. {\bf 91}, 015005 (2019).






\bibitem{vanveen1}E. van Veen, S. Yuan, M.I. Katsnelson, M. Polini and M.Tomadin, Phys. Rev. B {\bf 93}, 1115428 (2016).
\bibitem{vanveen2}E. van Veen, A. Tomadin, M. Polini, M.I. Katsnelson and S. Yuan, Phys. Rev. B, {\bf 96}, 235438 (2017).
\bibitem{fremling}M. Fremling, M. van Hooft, C. Morais Smith, and L. Fritz, Phys. Rev. Research {\bf 2}, 013044 (2020).

\bibitem{westerhout} T. Wasterhout, E. van Veen, M.I. Katsnelson and S. Yuan, Phys. Rev. B {\bf 97}, 205434 (2018).
\bibitem{brzezinska} M. Brzezinska, A. M. Cook, and T. Neupert, Phys. Rev. B {\bf 98}, 205116 (2018).





\bibitem{bouzerar-mayou} G. Bouzerar and D. Mayou, accepted for publication in Phys. Rev. Research.




\bibitem{das-sarma} S. Das Sarma, S. Adam, E. H. Hwang and E. Rossi, Rev. Mod. Phys. {\bf 83}, 407 (2011).
\bibitem{castro-neto} A.H. Castro Neto, F. Guinea, N. M. R. Peres, K. S. Novoselov, and A. K. Geim, Rev. Mod. Phys. {\bf 81}, 109 (2009).
\bibitem{geim} A. K. Geim, and K. S. Novoselov, Nature Mater. {\bf 6}, 183 (2007).
\bibitem{falko} Fal'ko V., A. Geim, S. Das Sarma, A. MacDonald, and P. Kim, Solid State Commun. {\bf 149}, 1039 (2009).





\bibitem{topology} "Topology and condensed matter physics", Editors: S.M. Bhattacharjee, et al., Springer Verlag (2017).
%\bibitem{balka} R. Balka, Z. Buczolich, M. Elekes, Adv. Math., \textbf{274}, 881 (2015).


%\bibitem{coey}M. Venkatesan, C. B. Fitzgerald and J. M. D. Coey, Nature \textbf{430}, 630 (2004)
%\bibitem{bouzerard0} G. Bouzerar and T. Ziman Phys. Rev. Lett. \textbf{96,} 207602  (2006).






\bibitem{shen} R. Shen, L. B. Shao, B. Wang, and D. Y. Xing, Phys. Rev. B \textbf{81}, 041410 (2010).
\bibitem{goldman} N. Goldman, D. F. Urban, and D. Bercioux, Phys. Rev. A \textbf{83}, 063601 (2011).
\bibitem{apaja} V. Apaja, M. Hyrkas, and M. Manninen, Phys. Rev. A \textbf{ 82}, 041402 (2010).
\bibitem{vicencio} R.A. Vicencio et al., Phys Rev. Lett. \textbf{114}, 245503 (2015).
\bibitem{guzman}  D Guzman-Silva, et al., New J. Phys. \textbf{16,} 063061 (2014).
\bibitem{mukherjee} S. Mukherjee et al., Phys. Rev. Lett. \textbf{114}, 245504 (2015).
\bibitem{cui} B. Cui et al. Nature Communications \textbf{11}, 66 (2020).


\bibitem{mucciolo} A. Ferreira and E. R. Mucciolo, Phys. Rev. Lett. {\bf 115}, 106601 (2015). 
\bibitem{weisse} A. Weisse, G. Wellein, A. Alvermann and H. Fehske, Rev. Mod. Phys. {\bf 78}, 275 (2006).


\bibitem{richard} R. Bouzerar et al., Phys. Rev. B  {\bf 94}, 094437 (2016).
\bibitem{lee} H. Lee, E.R. Mucciolo, G. Bouzerar and S. Kettemann, Phys. Rev. B  {\bf 86}, 205427 (2012).



\bibitem{cresti} A. Cresti, F. Ortmann, T. Louvet, D. Van Tuan and S. Roche, Phys. Rev. Lett. {\bf 110}, 196601 (2013).
\bibitem{trambly} G. Trambly de Laissardiere and D. Mayou, Phys. Rev. Lett. {\bf 111},146601 (2013).
\bibitem{triozon} F. Triozon, J. Vidal, R. Mosseri and D. Mayou, Phys. Rev. B  {\bf 65}, 220202 (2002).

\bibitem{pereirab} V.M. Pereira, J.M.B. Lopes dos Santos and A.H. Castro, Phys. Rev. B {\bf 77}, 115109 (2008).

\bibitem{lieb} E.H. Lieb Phys. Rev. Lett., {\bf 62}, 1201 (1989).

%\bibitem{SM} G. Bouzerar and D. Mayou, Supplementary material.


\bibitem{trambly2} G. Trambly de Laissardiere J.-P. Julien and D. Mayou, Phys. Rev. Lett., \textbf{97}, 026601 (2006).


\bibitem{garcia} J. H. Garcia, L. Covaci, and T. G. Rappoport, Phys. Rev. Lett. \textbf{114}, 116602 (2015).

\bibitem{kubo} R. Kubo, J. Phys. Soc.  Japan, \textbf{12}, 570 (1957).

\bibitem{greenwood}  D.A. Greenwood, Proceedings of the Physical Society, \textbf{71}, 585 (1958).



\end{thebibliography}
\end{document}